\documentclass[aps,prd,twocolumn,amsmath]{revtex4-2}

\usepackage{graphicx}
\usepackage{float}
\usepackage{epstopdf}
\usepackage{epsfig}
\usepackage{amsmath}
\usepackage{subcaption}%\usepackage{subfig}
\usepackage{color}
\usepackage{amssymb}   % \gtrsim, \geqslant, etc etc: see amsguide.ps
\usepackage{amsfonts}  % \mathfrak and \mathbb{x} (Blackboard bold)
\usepackage{amsbsy}    % \pmb and \boldsymbol
\usepackage{pstricks}
\usepackage{slashed}
\usepackage{dcolumn}
\usepackage{xparse}
\usepackage{bm}
\usepackage{hyperref}
\usepackage{soul}
\usepackage{caption}

\newcommand{\al}{\alpha}

\newcommand{\beq}{\begin{equation}}
\newcommand{\eeq}{\end{equation}}
\newcommand{\ba}{\begin{array}}
\newcommand{\ea}{\end{array}}
\newcommand{\bea}{\begin{align}}
\newcommand{\eea}{\end{align}}
\newcommand{\bi}{\begin{itemize}}
\newcommand{\ei}{\end{itemize}}
\newcommand{\ben}{\begin{enumerate}}
\newcommand{\een}{\end{enumerate}}
\newcommand{\bc}{\begin{center}}
\newcommand{\ec}{\end{center}}
\newcommand{\bl}{\begin{flushleft}}
\newcommand{\el}{\end{flushleft}}
\newcommand{\br}{\begin{flushright}}
\newcommand{\er}{\end{flushright}}

\newcommand{\nn}{\nonumber \\}
\newcommand\Eqn[1]{Eq.~(\ref{#1})}  % includes ``Eq.'' in front
\newcommand\Fig[1]{Fig.~\ref{#1}} % includes ``Fig.'' in front

\newcommand{\ie}{{i.e.}}

\newcommand{\GeV}{{\rm GeV}}

\newcommand\comment[1]{ \hbox{[{\it Comment suppressed here.}\/]} }
\newcommand\hide[1]{}
\newcommand{\skipover}[1]{}

\DeclareUnicodeCharacter{03D6}{\pisymbol} 
\begin{document}
%\title{Symmetry Preserving Regularization in Contact Model: taking kaon poreperties for example}
%\title{A Symmetry Preserving Contact Interaction: taking kaon properties for example}
\title{A symmetry preserving contact interaction treatment of the kaon}
\author{Zanbin Xing}\email{xingzb@mail.nankai.edu.cn}
\author{Lei Chang}\email{leichang@nankai.edu.cn}
\affiliation{School of Physics, Nankai University, Tianjin 300071, China}
%
%\date{\today}
%
\begin{abstract}
A symmetry-preserving regularization procedure for dealing with the contact interaction model is proposed in this work. This regularization procedure follows a series of consistency conditions which are necessary to maintain gauge symmetry. Under this regularization, proofs for the preservation of the Ward-Takahashi identities are given and the loop integrals in the contact interaction model are systematically computed. As an application example, the kaon electromagnetic form factor and $K_{l3}$ transition form factor are computed and self-consistent results are obtained. Since the proposed regularization properly handles the divergence, one is freed from the inconsistencies caused by the regularization and can concentrate more on the physical discussion.
\end{abstract}
\maketitle
\section{introduction}
Over the years, the Dyson-Schwinger Equations(DSEs) have proved to be a powerful tool for studying non-pertubative nature of Quantum Chromodynamics(QCD) in the continuum\cite{Roberts:1994dr,Maris:2003vk,Fischer:2018sdj}. Self-consistent treatments of the quark gap equation and bound state equations, such as Bethe-Salpeter equation(BSE) and Faddeev equation, creates a bridge connecting the hadrons to the fundamental degrees of freedom of QCD, quarks and gluons\cite{Eichmann:2009zx,Eichmann:2016yit}. 

Within the framework of DSEs, a vector-vector contact interaction approximation was proposed in Ref.~\cite{Gutierrez-Guerrero:2010waf}. Despite its simplicity of the contact interaction model in describing the real world, it has been used to calculate a wide range of hadron properties, including mass spectrum, various decay processes, electromagnetic form factors and transition form factors, and the parton distributions, see Ref.~\cite{Gutierrez-Guerrero:2010waf,Roberts:2011cf,Roberts:2011wy,Chen:2012qr,Chen:2012txa,Segovia:2013uga,Segovia:2014aza,Xu:2015kta,Bedolla:2015mpa,Serna:2016kdb,Lu:2017cln,Serna:2017nlr,Gutierrez-Guerrero:2019uwa,Zhang:2020ecj,Gutierrez-Guerrero:2021rsx,Xing:2021dwe}. However, due to the non-renormalizable nature of the contact interaction, the regularization scheme becomes a crucial part in the practical calculation and a good regularization scheme should properly characterise the divergence structure of the theory. It is worth noting that symmetries, and in particular the Ward-Takahashi identities(WTIs), provide a strong constraint that must be preserved during the regularization process. 

The regularization procedures performed in previous studies of contact model have been unsatisfactory. Firstly, WTIs do not naturally hold under the previous regularization procedure. Moreover, there are cases where inconsistent results occur in the calculation of kaon electromagnetic form factor, when the principle of charge conservation is violated~\cite{Chen:2012txa}. The main reason for these problems is that previous regularization procedure fails to properly deal with the quadratic and logarithmic divergent integrals.

Inspired by Ref.~\cite{Wu:2002xa} we presented a new regularization procedure in this work. This proper regularization meets many interesting properties of the dimensional regularization without changing the space-time dimension. One of the most fascinating properties is that gauge symmetries are preserved under this regularization.

This paper is organized as follows, Sec.~\ref{sec::regularization} introduces the new symmetry preserving regularization that properly handle the divergent integrals. Sec.~\ref{sec::contactmodel} discusses this new regularization in the contact interaction model, incorporating the preserving of (axial-)vector WTIs, and gives steps for the systematic calculations with contact model.  Sec.~\ref{sec::kaon} provides results of the kaon electromagnetic form factor and $K_{l3}$ form factor under the new regularizaiton, and the last section gives a brief summary.

\section{symmetry preserving regularization}\label{sec::regularization}
%Before a detailed regularization, 
Before discussing the regularization procedures in detail, it is helpful to introduce the so-called one-fold irreducible loop integrals(ILIs) in Ref.~\cite{Wu:2002xa}.
\begin{eqnarray}
I_{-2\alpha}(\mathcal{M}^2)&=&\int_{q}\frac{1}{(q^2+\mathcal{M}^2)^{\alpha+2}}\,,\nn
I^{\mu\nu}_{-2\alpha}(\mathcal{M}^2)&=&\int_{q}\frac{q_{\mu}q_{\nu}}{(q^2+\mathcal{M}^2)^{\alpha+3}}\,,\nn
I^{\mu\nu\rho\sigma}_{-2\alpha}(\mathcal{M}^2)&=&\int_{q}\frac{q_{\mu}q_{\nu}q_{\rho}q_{\sigma}}{(q^2+\mathcal{M}^2)^{\alpha+4}}
\end{eqnarray}
with $\int_q\doteq\int\frac{d^4q}{(2\pi)^4}$ and $\alpha=-1,0,1,\cdots$. Here $\alpha=-1$ represents quadratically divergent integrals and $\alpha=0$ represents logarithmically divergent integrals. With the help of Feynman parametrezation, it is straightforward to conclude that all one loop integrals can be expressed in terms of these integrals. Where $\mathcal{M}$ is a function of Feynman parameters, external momenta and the corresponding mass scales. A regularization procedure can be implemented after rearranging one loop integrals into these ILIs. 

In Ref.~\cite{Wu:2002xa}, a loop regularization is proposed which simulates in many interesting features to the momentum cutoff, Pauli-Villars and dimensional regularization without modifying the original Lagrangian formalism, and is directly performed in the space-time dimension of original theory. The loop regularization is equivalent to introducing a weight function to regularize the proper-time variable $\tau$ integration\cite{Wu:2003dd},for example,
\begin{eqnarray}
I^{LR}_{-2\alpha}(\mathcal{M}^2)=\lim_{N \to \infty}\int_{q}\int_{0}^{\infty}& &d\tau\mathcal{W}_N(\tau,M_c,\mu_s)\nn
& &\times\frac{\tau^{\alpha+1}}{\Gamma(\alpha+2)}e^{-\tau(q^2+\mathcal{M}^2)}\,,
\end{eqnarray}
where the superscript $LR$ denotes the loop regularization and $\Gamma(n)$ is the gamma function. An explicit form of the weight function is
\begin{equation}
\mathcal{W}_N(\tau,M_c,\mu_s)=e^{-\tau \mu_s^2}(1-e^{-\tau M_R^2})^N\,,
\end{equation}
with $M_R^2=M_c^2 h_w(N) lnN$, $h_w(N)\gtrsim1$ and $h_w(N\rightarrow\infty)=1$. The two energy scale $M_c$ and $\mu_s$ serve as ultraviolet(UV) and infrared(IR) cutoff, respectively. It is worth noting that when $N\rightarrow\infty$ and $\mu_s=0$ the weight function becomes
\begin{equation}
\lim_{N \to \infty}\mathcal{W}_N(\tau,M_c,\mu_s=0)=\theta(\tau M_c^2-1)\,.
\end{equation}
Thus the weight function regularizes the proper-time integral just as regularizes it with a hard UV cutoff $1/M_c^2$.

We introduce a regularization procedure which is based on the Schwinger's proper-time method. The regularization procedure for the scalar type ILIs is
\begin{eqnarray}
I_{-2\alpha}(\mathcal{M}^2)&=&\int_{q}\frac{1}{(q^2+\mathcal{M}^2)^{\alpha+2}}\nn
                    &=&\int_{q}\int_{0}^{\infty}d\tau \frac{\tau^{\alpha+1}}{\Gamma(\alpha+2)}e^{-\tau(q^2+\mathcal{M}^2)}\nn
                    &=&\int_{0}^{\infty}d\tau \frac{\tau^{\alpha-1}}{\Gamma(\alpha+2)}\frac{e^{-\tau\mathcal{M}^2}}{16\pi^2}\nn
\label{eqn::sreg}
\rightarrow I_{-2\alpha R}(\mathcal{M}^2)&=&\int_{\tau_{uv}^2}^{\tau_{ir}^2}d\tau \frac{\tau^{\alpha-1}}{\Gamma(\alpha+2)}\frac{e^{-\tau\mathcal{M}^2}}{16\pi^2}\,.
\end{eqnarray}
The label $R$ in the subscript denotes the regularized integrals. It is already seen that a hard UV cutoff $\tau_{uv}=1/M_c$ is equivalent to the loop regularization with $\mu_s=0$. However, instead of the sliding energy scale $\mu_s$ in the loop regularization, we introduce an hard IR cutoff $\tau_{ir}$ to implement confinement, as proposed in Ref.~\cite{Ebert:1996vx}. This way of dealing with the IR cutoff matches the regulators in previous contact model studies, which, as we shall see, can also maintain gauge symmetries if the tensor type ILIs are properly regularized.
Before proceeding, it is noted that when integer $\alpha<-1$, the loop integral vanishes under \Eqn{eqn::sreg}, which happens to be the same property of dimensional regularization.

Turning now to the regularization of the tensor type ILIs, the regularization procedure is
\begin{eqnarray}
I^{\mu\nu}_{-2\alpha}(\mathcal{M}^2)&=&\int_{q}\frac{q_{\mu}q_{\nu}}{(q^2+\mathcal{M}^2)^{\alpha+3}}\nn
             &=&\int_{q}\int_{0}^{\infty}d\tau q_{\mu}q_{\nu}\frac{\tau^{\alpha+2}}{\Gamma(\alpha+3)} e^{-\tau(q^2+\mathcal{M}^2)}\nn
             &=&\int_{q}\int_{0}^{\infty}d\tau \delta_{\mu\nu}\frac{q^2}{4}\frac{\tau^{\alpha+2}}{\Gamma(\alpha+3)} e^{-\tau(q^2+\mathcal{M}^2)}\nn
             &=&\delta_{\mu\nu}\int_{0}^{\infty}d\tau \frac{\tau^{\alpha-1}}{\Gamma(\alpha+3)}\frac{e^{-\tau\mathcal{M}^2}}{32\pi^2}\nn
\label{eqn::tr}
\rightarrow I^{\mu\nu}_{-2\alpha R}(\mathcal{M}^2)&=&\delta_{\mu\nu}\int_{\tau_{uv}^2}^{\tau_{ir}^2}d\tau \frac{\tau^{\alpha-1}}{\Gamma(\alpha+3)}\frac{e^{-\tau\mathcal{M}^2}}{32\pi^2}\,,
\end{eqnarray}
and
\begin{eqnarray}
I^{\mu\nu\rho\sigma}_{-2\alpha}(\mathcal{M}^2)&=&\int_{q}\frac{q_{\mu}q_{\nu}q_{\rho}q_{\sigma}}{(q^2+\mathcal{M}^2)^{\alpha+4}}\nn
             &=&\int_{q}\int_{0}^{\infty}d\tau q_{\mu}q_{\nu}q_{\rho}q_{\sigma}\frac{\tau^{\alpha+3}}{\Gamma(\alpha+4)}e^{-\tau(q^2+\mathcal{M}^2)}\nn
             &=&\int_{q}\int_{0}^{\infty}d\tau S_{\mu\nu\rho\sigma}\frac{q^4}{24} \frac{\tau^{\alpha+3}}{\Gamma(\alpha+4)}e^{-\tau(q^2+\mathcal{M}^2)}\nn
             &=&S_{\mu\nu\rho\sigma}\int_{0}^{\infty}d\tau \frac{\tau^{\alpha-1}}{\Gamma(\alpha+4)}\frac{e^{-\tau\mathcal{M}^2}}{64\pi^2}\nn
\rightarrow I^{\mu\nu\rho\sigma}_{-2\alpha R}(\mathcal{M}^2)&=&S_{\mu\nu\rho\sigma}\int_{\tau_{uv}^2}^{\tau_{ir}^2}d\tau \frac{\tau^{\alpha-1}}{\Gamma(\alpha+4)}\frac{e^{-\tau\mathcal{M}^2}}{64\pi^2}\,,
\end{eqnarray}
where $S_{\mu\nu\rho\sigma}=\delta_{\mu\nu}\delta_{\rho\sigma}+\delta_{\mu\rho}\delta_{\sigma\nu}+\delta_{\mu\sigma}\delta_{\nu\rho}$ is the total symmetric tensor. It is obvious that the regularized tensor type ILIs and scalar type ILIs are related as follows
\begin{eqnarray}
I^{\mu\nu}_{-2\alpha R}(\mathcal{M}^2)&=&\frac{\Gamma(\alpha+2)}{2\Gamma(\alpha+3)}\delta_{\mu\nu}I_{-2\alpha R}(\mathcal{M}^2)\,,\\
I^{\mu\nu\rho\sigma}_{-2\alpha R}(\mathcal{M}^2)&=&\frac{\Gamma(\alpha+2)}{4\Gamma(\alpha+4)}S_{\mu\nu\rho\sigma}I_{-2\alpha R}(\mathcal{M}^2)\,.
\end{eqnarray}
These relations are precisely the so-called consistency conditions of gauge symmetry in Ref.~\cite{Wu:2002xa,Wu:2003dd}, which are independent of regularization and are necessary for preserving the gauge invariance of theories. It is noted that ILIs under dimensional regularization also satisfy these conditions. These consistency conditions connect tensor type ILIs and scalar type ILIs and then any gauge invariant theories can be properly described in terms of the regularized scalar type ILIs. In fact, there are a series of consistency conditions for ILIs with even more Lorentz index which are rarely to encountered and are therefore not presented here.

We now consider the regularization procedure in previous contact model studies, such as in Ref.~\cite{Gutierrez-Guerrero:2010waf}. For example, the quadratic divergent tensor type ILI is regularized as follows,
\begin{eqnarray}
I^{\mu\nu}_{2}(\mathcal{M}^2)&=&\int_{q}\frac{q^{\mu}q^{\nu}}{(q^2+\mathcal{M}^2)^2}\nn
             &\rightarrow&\int_{q}\frac{\delta_{\mu\nu}}{4}\left(\frac{1}{q^2+\mathcal{M}^2}-\frac{\mathcal{M}^2}{(q^2+\mathcal{M}^2)^2}\right)\nn
             &=&\int_{q}\int_{0}^{\infty}d\tau \frac{\delta_{\mu\nu}}{4}e^{-\tau(q^2+\mathcal{M}^2)}(1-\tau\mathcal{M}^2)\nn
             &=&\delta_{\mu\nu}\int_{0}^{\infty}d\tau \frac{e^{-\tau\mathcal{M}^2}}{64\pi^2\tau^2}(1-\tau\mathcal{M}^2)\nn
\label{eqn::trp}
\rightarrow I^{\mu\nu}_{2R'}(\mathcal{M}^2)&=&\delta_{\mu\nu}\int_{\tau_{uv}^2}^{\tau_{ir}^2}d\tau \frac{e^{-\tau\mathcal{M}^2}}{64\pi^2\tau^2}(1-\tau\mathcal{M}^2)\,,
\end{eqnarray}
where $R'$ denotes the regularization in previous contact model studies. Apparently, this way of regularization does not satisfy the consistency conditions and hence breaks the gauge symmetries, especially the WTIs. This inappropriate  regularization of the tensor type ILIs is also responsible for the inconsistent results that appear in the more complex computations as the case of the triangle diagrams.
The key difference between the two regularizations $R$ and $R'$ is which step to do the symmetry analysis, \ie, the substitution like
\begin{eqnarray}
q_{\mu}q_{\nu}\rightarrow \delta_{\mu\nu}\frac{q^2}{4}\,.
\end{eqnarray}
In \Eqn{eqn::tr}, one first perform the proper-time method so that the divergent momentum integrals become well-defined, followed by the substitution. In this way, the tensor type integrals are properly regularized and gauge symmetry is preserved. On the contrary, in \Eqn{eqn::trp} the substitution is made while the integrals are still divergent, then the divergent structures of theories are destroyed and leads to possible inconsistency in further calculations. It is well known that the substitution is only valid when the loop integral is convergent. In principle, if we push the regulators $\tau_{uv}\to 0$ and $\tau_{ir}\to \infty$, which restores the original ILIs, the two regularizations $R$ and $R'$ lead to the same results for convergent integrals. However, we keep the two regulators $\tau_{uv}$ and $\tau_{ir}$ even when the ILIs are convergent, so that the regularized ILIs still retain the following relation to the original ILIs
\begin{equation}
\label{eqn::derivative}
I_{-2(\alpha+1)R}(\mathcal{M}^2)=-\frac{1}{\alpha+2}\frac{d}{d\mathcal{M}^2}I_{-2\alpha R}(\mathcal{M}^2)\,.
\end{equation}
In the following sections, we will adopt the new regularization $R$ to illustrate how these consistency conditions keep gauge symmetries and then recalculate the kaon electromagnetic and transition form factors. For simplicity, the label $R$ is suppressed thereafter.

\section{symmetry preserving regularization in the contact model}\label{sec::contactmodel}
The quark gap equation in the contact model can be written as
\begin{equation}
S^{-1}_{f}(p)=S_{f0}^{-1}(p)+\frac{4}{3m_g^2}\int_q{\gamma_{\al}S_{f}(q)\gamma_{\al}}\,,
\end{equation}
where $S^{-1}_{f}(p)$ is the inverse of the dressed quark propagator with flavor $f$ and momentum $p$, which has the general form
\begin{equation}
S^{-1}_{f}(p)=i\slashed{p}+M_f\,,
\end{equation}
where $M_f$ is independent of momentum due to the features of contact interaction.
$S^{-1}_{f0}=i\slashed{p}+m_f$ is the inverse of the bare quark propagator with $m_f$ being the current quark mass.

In terms of the ILIs, the gap equation becomes
\begin{equation}
\hat{M}_f=\frac{16M_f}{3m_g^2}I_2(M_f^2),
\end{equation}
where $\hat{M}_f=M_f-m_f$.

In the contact model, the meson's Bethe-Salpeter amplitude(BSA) depends only on meson's total momentum Q, thus the most general form of the pseudoscalar(PS) meson BSA with a outgoing $a$ quark and an incoming $b$ quark can be expressed as
\begin{equation}
\Gamma^{ab}_{PS}(Q)=i\gamma_5 E_{PS}(Q)+\frac{\gamma_5\slashed{Q}}{M_{ab}}F_{PS}(Q)\,,
\end{equation}
where we have introduced the usual reduced mass $M_{ab}=\frac{2M_aM_b}{M_a+M_b}$. The BSAs satisfy the following homogeneous Bethe-Salpeter equation
\begin{eqnarray}\label{eqn::mbse}
\Gamma^{ab}_{PS}(Q)&=&-\frac{4}{3m_{G}^{2}}\int_{q}\gamma_{\al}\chi^{ab}_{PS}(q,q_{-})\gamma_{\al}\,,
\end{eqnarray}
where $q_{-}=q-Q$ and $\chi^{ab}_{PS}(q,q_{-})=S_{a}(q)\Gamma^{ab}_{PS}(Q)S_{b}(q_{-})$ is the corresponding Bethe-Salpeter wave function.

In the following, we will show how the new regularization procedure preserves the axial-vector WTI exactly. The axial-vector WTIs states
\begin{eqnarray}\label{eqn::avwti}
Q_{\mu}\Gamma^{ab}_{5\mu}(Q)=S^{-1}_{a}(q)i\gamma_5&+&i\gamma_5 S^{-1}_{b}(q_{-})\nn
                                        &-&i(m_a+m_b)\Gamma^{ab}_{5}(Q)\,.
\end{eqnarray}
The axial-vector vertex $\Gamma^{ab}_{5\mu}(Q)$ satisfies an inhomogeneous BSE
\begin{equation}\label{eqn::avbse}
\Gamma^{ab}_{5\mu}(Q)=\gamma_5\gamma_\mu-\frac{4}{3m_{G}^{2}}\int_q\gamma_{\al}\chi^{ab}_{5\mu}(Q)\gamma_{\al}\,,
\end{equation}
with $\chi^{ab}_{5\mu}(Q)=S_{a}(q)\Gamma^{ab}_{5\mu}(Q)S_{b}(q_{-})$. The pseudoscalar vertex $\Gamma_{5}(Q)$ satisfies the equation in analogy with the axial-vector vertex by replacing inhomogeneous term $\gamma_5\gamma_\mu$ with $\gamma_5$.
The axial-vector WTI connects the vertex BSEs and the gap equation, leading to the following two identities
\begin{eqnarray}
\hat{M}_a+\hat{M}_b=\frac{4}{3m_{G}^{2}}\int_q\left(\frac{4M_a}{q^2+M_a^2}+\frac{4M_b}{q_{-}^2+M_b^2}\right)\,,\nn
0=\frac{4}{3m_{G}^{2}}\int_q\left(\frac{2Q\cdot q}{q^2+M_a^2}-\frac{2Q\cdot q_{-}}{q_{-}^2+M_b^2}\right)\,.
\end{eqnarray}
By analysing the integrals with Feynman parametrization, one can rearrange the expressions in terms of ILIs and arrive at
\begin{eqnarray}\label{eqn::WTI1}
\hat{M}_a+\hat{M}_b&=&\frac{16(M_a+M_b)}{3m_{G}^{2}}\int_{0}^{1}du \{I_2(\omega_2)+\frac{(uM_b-\bar{u}M_a)}{M_a+M_b}\nn
& &\times I_0(\omega_2)((2u-1)Q^2-M_b^2+M_a^2)\}\,,
\end{eqnarray}
and
\begin{equation}\label{eqn::WTI2}
0=\frac{8}{3m_{G}^{2}}\int_{0}^{1}du\{Q^2I_{2}(\omega_2)-2Q_{\mu}Q_{\nu}I_2^{\mu\nu}(\omega_2)\}\,,
\end{equation}
where $\bar{u}=1-u$ and $\omega_2=u M_b^2+\bar{u} M_a^2+u\bar{u}Q^2$.

%For the first equation, lets define
Consider the following integral
\begin{equation}\label{eqn::hav}
H_{AV}=\frac{16(M_a+M_b)}{3m_{G}^{2}}\int_{0}^{1} g_{AV}(u,M_a,M_b,Q^2)du\,,
\end{equation}
with
\begin{equation}
g_{AV}(u,M_a,M_b,Q^2)=\frac{d}{du}\left[I_2(\omega_2)\frac{(uM_b-\bar{u}M_a)}{M_a+M_b}\right]\,.
\end{equation}
On the one hand, one can evaluate this integral directly
\begin{eqnarray}
H_{AV}&=&\frac{16(M_a+M_b)}{3m_{G}^{2}}\left.I_2(\omega_2)\frac{(uM_b-\bar{u}M_a)}{M_a+M_b}\right|^{1}_{0}\nn
 &=&\frac{16}{3m_{G}^{2}}(M_a I_2(M_a^2)+M_b I_2(M_b^2))\,.
\end{eqnarray}
On the other hand, one can split the total derivative before integration, and it is easy to see that(with the help of the properties in \Eqn{eqn::derivative})
\begin{eqnarray}
H_{AV}&=&\frac{16(M_a+M_b)}{3m_{G}^{2}}\int_{0}^{1}du \{I_2(\omega_2)+\frac{(uM_b-\bar{u}M_a)}{M_a+M_b}\nn
& &\times I_0(\omega_2)((2u-1)Q^2-M_b^2+M_a^2)\}\,,
\end{eqnarray}
which is precisely the right hand side \Eqn{eqn::WTI1}. Then the \Eqn{eqn::WTI1}is nothing but the sum of two quark gap equations.

The right hand side of \Eqn{eqn::WTI2} is exactly zero due to the consistent conditions contained in the new regularization $R$. While in Ref.~\cite{Gutierrez-Guerrero:2010waf}, the vanish of the right hand side of \Eqn{eqn::WTI2} is imposed by hand. The reason is that the previous regularization $R'$ fails to fulfill the consistency conditions and thus gauge symmetry is explicitly broken.

Let's now focus on the vector vertex which satisfies the vector WTI
\begin{eqnarray}\label{eqn::vwti}
Q_{\mu}i\Gamma^{ab}_{\mu}(Q)=S^{-1}_{a}(q)&-&S^{-1}_{b}(q_{-})\nn
                                          &-&(m_a-m_b)\Gamma^{ab}_{I}(Q)\,,
\end{eqnarray}
and follows the inhomogeneous BSE
\begin{eqnarray}
\Gamma^{ab}_{\mu}(Q)=\gamma_{\mu}-\frac{4}{3m_{G}^{2}}\int_{q}\gamma_{\al}\chi^{ab}_{\mu}(Q)\gamma_{\al}\,,
\end{eqnarray}
with $\chi^{ab}_{\mu}(Q)=S^{a}(q)\Gamma^{ab}_{\mu}(Q)S^{b}(q_{-})$. The scalar vertex $\Gamma^{ab}_{I}(Q)$ follows a similar equation with inhomogeneous term being the identity matrix $I_{D}$. Under the contact interaction , the general form of the vector and scalar vertexes can be written as
\begin{eqnarray}
\Gamma^{ab}_{\mu}(Q)&=&\gamma_{\mu}^{L} V^{ab}_{L}(Q^2)+\gamma_{\mu}^{T} V^{ab}_{T}(Q^2)\nn
&&+ I_{D}(-i Q_{\mu})V^{ab}_{S}(Q^2)\,,\\
\Gamma^{ab}_I(Q)&=&I_{D} S^{ab}_{1}(Q^2)-\frac{i\slashed{Q}}{M_{ab}}S^{ab}_{2}(Q^2) \,,
\end{eqnarray}
where $\gamma_\mu^L=\frac{Q_\mu\slashed{Q}}{Q^2}$, $\gamma_\mu^T=\gamma_\mu-\gamma_\mu^L$. In Ref.~\cite{Xing:2021dwe}, an additional interaction kernel is proposed which maintains the WTIs. This additional kernel can automatically generate a quark anomalous moment term $\sigma_{\mu\nu}Q_{\nu}$ in the vector vertex, which has significant impact on vector meson. However, since the additional kernel does not influence the discussion in this paper, we will not adopt it for simplicity.

The most straightforward way of solving the BSE equations is to compute the momentum integrals in the BSE directly. However, this method is computationally cumbersome when evaluating the $\gamma$ matrix. Thus we solve the vector and scalar vertex BSEs by projecting the BSE with the various projecting operators. However, it is stressed that the Lorentz index of vector or tensor objects is important and should be retained before regularization. As an example, we illustrate this requirement through the solution of vector vertex,
\begin{eqnarray}
V^{ab}_T(Q^2)=\frac{1}{1-f^{ab}_{T}(Q^2)}\,,
\end{eqnarray}
with
\begin{equation}
f^{ab}_{T}(Q^2)=-\frac{4}{3m_{G}^{2}}\text{tr}_{D}\delta_{\mu\nu}\int_q P^T_\nu \gamma_{\al}S^{a}(q)\gamma^T_\mu S^{b}(q_{-})\gamma_{\al}\,,
\end{equation}
by using the projection operator $P^T_\nu=\frac{\gamma^T_\nu}{\text{tr}_{D}(\gamma^T_\mu\gamma^T_\mu)}$. Here we express 
\begin{equation}\label{eqn::index}
P^T_\mu\otimes\gamma^T_\mu \to \delta_{\mu\nu}P^T_\nu\otimes\gamma^T_\mu\,,
\end{equation}
which is used to retain the Lorentz index of the vector vertex $\Gamma^{ab}_{\mu}(Q)$ herein. Analyzing the integral with Feynman parametrization and rearranging it into ILIs, and then regularizing the ILIs with $\delta_{\mu\nu}$ uncontracted, yields the following expression
\begin{eqnarray}
f^{ab}_{T}(Q^2)&=&\frac{4}{3m_{G}^{2}}\delta_{\mu\nu}\int_{0}^{1}du\frac{2}{3}(\frac{Q_\mu Q_\nu}{Q^2}-\delta_{\mu\nu})I_0(\omega_2)\nn
       & &\times (u M_b^2-M_b M_a+\bar{u}M_a^2+2u\bar{u}Q^2)\,.
\end{eqnarray}
It can be seen that the transverse structure is exactly maintained with the index of vector vertex uncontracted before regularization. If the contraction is performed before regularization, the transverse structure may be destroyed. 

By solving the corresponding BSE, the rest of the dressing functions in these two vertices are
\begin{eqnarray}
V^{ab}_{L}(Q^2)&=&\frac{1+4f^{ab}_1(Q^2)}{1-f^{ab}_{L}(Q^2)}\,,\\
V^{ab}_{S}(Q^2)&=&\frac{-4f^{ab}_2(Q^2)}{1-f^{ab}_{L}(Q^2)}\,,\\
S^{ab}_{1}(Q^2)&=&\frac{1+2(M_b-M_a)f^{ab}_2(Q^2)}{1-f^{ab}_{L}(Q^2)}\,,\\
S^{ab}_{2}(Q^2)&=&\frac{2M_{ab}f^{ab}_2(Q^2)}{1-f^{ab}_{L}(Q^2)}\,,
\end{eqnarray}
where
\begin{eqnarray}
f^{ab}_{L}(Q^2)&=&8Q^2(f^{ab}_2(Q^2))^2-2(M_b-M_a)f^{ab}_2(Q^2)\nn
          & &-4f^{ab}_1(Q^2)(1+2(M_b-M_a)f^{ab}_2(Q^2))\,,\\
f^{ab}_1(Q^2)&=&\frac{4}{3m_g^2}\int_{0}^{1}[I_0(\omega_{2})(u M_b^2+\bar{u}M_a^2\nn
& &+M_b M_a+2u\bar{u}Q^2)-I_2(\omega_{2})]\,,\\
f^{ab}_2(Q^2)&=&\frac{4}{3m_g^2}\int_{0}^{1}I_0(\omega_{2})(u M_b-\bar{u}M_a)\,.
\end{eqnarray}
Plugging the solved vector and scalar vertices into the vector WTI \Eqn{eqn::vwti}, one finds the following equation
\begin{equation}\label{eqn::gapminus}
\hat{M}_b-\hat{M}_a=-4(M_b-M_a)f^{ab}_1(Q^2)+4Q^2f^{ab}_2(Q^2)\,.
\end{equation}
This equation can be proved in analogy with \Eqn{eqn::hav} by introducing
\begin{equation}
H_{V}=\frac{16}{3m_{G}^{2}}\int_{0}^{1} g_{V}(u,M_a,M_b,Q^2)du\,,
\end{equation}
where
\begin{equation}
g_{V}(u,M_a,M_b,P^2)=\frac{d}{du}\left[I_2(\omega_2)(u M_b+\bar{u}M_a)\right]\,.
\end{equation}
One finds \Eqn{eqn::gapminus} is exactly the difference between the two quark gap equations.

In the end of this section, we list the systematic procedures in regularizing the 1-loop integrals in the contact model.
\begin{itemize}
\item[1.]~First, one uses Feynman parametrization to analyze the integral and then rearranges the expressions in terms of the ILIs;

\item[2.]~Regularize the ILIs through the new regularization and express all regularized tensor type ILIs in terms of scalar type ones through the consistent conditions;

\item[3.]~When evaluating formulae that contain vector or tensor objects, their Lorentz index should be retained rather than contracted before the tensor type ILIs are regularized.

\end{itemize}
The first step is trivial, but provides a convenient way to manage the expressions in calculation. We can then easily implement regularization in step 2. We emphasis again a proper regularization should let the regularized tensor type ILIs satisfy the consistent conditions, otherwise gauge symmetries will be broken. The third rule concerns vector and tensor objects, whose Lorentz metrics are important. To be clear, this rule charges not only for solving the BSEs, but also for any physical quantity that includes vector and tensor objects. An easy way of implementing this rule can be achieved by, say, \Eqn{eqn::index}, which will generate the following typical integrals
\begin{equation}
\delta_{\mu\nu}\int_{q}\frac{q_{\mu}q_{\nu}}{(q^2+\mathcal{M}^2)^{\alpha+2}}\,,
\end{equation}
where $\delta_{\mu\nu}$ comes from \Eqn{eqn::index}. It is apparent that different orders of regularization and contraction lead to different results, take $\alpha=-1$ for example:
\begin{eqnarray}
&&\delta_{\mu\nu}\int_{q}\frac{q_{\mu}q_{\nu}}{(q^2+\mathcal{M}^2)^{2}}\to\delta_{\mu\nu}I_{2R}^{\mu\nu}=2I_{2R}\,,\\
&&\delta_{\mu\nu}\int_{q}\frac{q_{\mu}q_{\nu}}{(q^2+\mathcal{M}^2)^{2}}= I_{2}-\mathcal{M}^2I_{0}\to I_{2R}-\mathcal{M}^2I_{0R}\,.\nn
\end{eqnarray}
Here arrows represent operations of regularization and equal signs represent operations of contraction. And we explicitly use the label $R$ to distinguish the regularized ILIs from the orignal ones. It is observed that if one contracts the index before regularization, the unregularized tensor ILIs are turned into the unregularized scalar ILIs, which incorrectly characterises the divergences of theories, and then some fundamental properties are lost, such as the current conservation.

\section{Kaon form factors}\label{sec::kaon}
As the simplest bound state with strangeness, kaon has attracted much attention since its discovery in the middle of the last century\cite{Rochester:1947mi} and has led to many important researches in standard model, such as the introduction of strangeness\cite{Gell-Mann:1953hzm}, the violation of parity\cite{Lee:1956qn}, quark mixing and the CKM matrix\cite{Cabibbo:1963yz,Kobayashi:1973fv}, CP violation\cite{Christenson:1964fg} and GIM mechanism\cite{Glashow:1970gm}. Recently, CERN has approved a world-unique QCD facility where highly intense and energetic kaon beams are used to map out the complete spectrum of excited kaons with an unprecedented precision\cite{Taboada-Nieto:2022igy}. This approval of CERN injects vitality into a wide range nowadays studies on low energy kaon phenomenology and high energy processes including kaon excitations, and opens a completely new horizon in kaon physics.

The kaon electromagnetic and transition form factor have also attracts fruitful studies\cite{Maris:2000sk,Xiao:2002iv,Stamen:2022uqh,Ji:2001pj,Bazavov:2012cd,RBCUKQCD:2015joy,Troitsky:2021vkw}. In previous contact model computation of these two form factors\cite{Chen:2012txa}, inconsistency results occurred due to the failure of the regularization to properly handle the divergent integrals.

To calculate the kaon electromagnetic form factor and the $K_{l3}$ transition form factor, three elements is required. The quark propagator and vector vertex are given in Sec.~\ref{sec::contactmodel}(in the following, $\Gamma_{\mu}^{ff}$ denotes the quark-photon vertex and $\Gamma_{\mu}^{su}$ denotes the quark–W boson vertex). The remaining ingredients are the pion and kaon BSAs, which are solved from the pseudoscalar BSEs. In solving this equation, we follow Ref.~\cite{Chen:2012txa}, the only difference being the regularization. Even though we get the same kernel as in Ref.~\cite{Chen:2012txa}, it is worth nothing that the following relation holds,
\begin{equation}
K_{FF}=-\frac{(M_s+M_u)^2}{2M_sM_u}K_{FE}\,.
\end{equation}
This is a natural result since the axial-vector WTI is automatically satisfied under the new regularization. Since the meson BSE is a homogeneous equation, normalization of the BSA is need for the computation of physical quantities. The canonical normalisation of pseudoscalar meson reads
\begin{equation}\label{eqn::norm}
Q_{\mu}=N_{c}\text{tr}_{D}\int_{q}\Gamma^{ab}_{PS}(-Q)S(q)\Gamma^{ab}_{PS}(Q)\frac{\partial}{\partial Q_{\mu}}S(q_{-})\,,
\end{equation}
and the leptonic decay constant reads
\begin{equation}\label{eqn::decay}
f^{ab}_{PS}Q_{\mu}=\frac{1}{\sqrt{2}}N_{c}\text{tr}_{D}\int_q\gamma_5\gamma_{\mu}\chi^{ab}_{PS}(q,q_{-})\,.
\end{equation}
The parameter are fixed to obtain the the following physical observables, $m_{u}=0.007GeV$, $m_{s}=0.17\GeV$, $m_{G}=0.132\GeV$, $\tau_{uv}=1/0.905\GeV^{-1}$, $\tau_{ir}=1/0.24\GeV^{-1}$.
\begin{table}[ht]
%\vspace*{-3mm}
\caption{\label{tab:massanddecay} masses and decay constants, the unit is \GeV}
\begin{tabular}{cccccc}
\hline
$M_u$ &$M_s$ &$m_{\pi}$ & $m_{K}$ & $f_{\pi}$ & $f_{K}$\\
\hline
$0.368$ &0.533 & 0.140 & 0.499 & 0.101 & 0.106\\
\hline
\end{tabular}
%\vspace*{-3mm}
\end{table}

\subsection{Kaon electromagnetic form factor}
In the impulse approximation, the kaon electromagnetic form factor is
\begin{equation}
F^{em}_{K^+}(Q^2)=e_u F^{u}_{K^+}(Q^2)+e_{s} F^{s}_{K^+}(Q^2)\,,
\end{equation}
where
\begin{eqnarray}
\label{eqn::fu}
2P_\mu F^{u}_{K^+}(Q^2)&=&N_c\text{tr}_{D}\int_q S^{s}(q)i\Gamma^{K^+}(-p)S^{u}(q+p)\nn
& &\times i\Gamma^{uu}_{\mu}(Q)S^{u}(q+k)i\Gamma^{K^+}(k)\,,\\
\label{eqn::fs}
2P_\mu F^{s}_{K^+}(Q^2)&=&N_c\text{tr}_{D}\int_q S^{s}(q-p)i\Gamma^{K^+}(-p)S^{u}(q)\nn
& &\times i\Gamma^{K^+}(k)S^{s}(q-k)i\Gamma^{ss}_{\mu}(Q)\,,
\end{eqnarray}
here $p=P+Q/2$, $k=P-Q/2$ with $k$ the incoming kaon momentum and $p$ the out going kaon momentum. The on-shell condition entails $P\cdot Q=0$ and $P^2=-m_K^2-Q^2/4$. We express the form factor formula as follows
\begin{eqnarray}\label{eqn::emff}
F^{u}_{K^+}(Q^2)&=&(F^{u}_{EE}(Q^2)E_{K^+}E_{K^+}+F^{u}_{EF}(Q^2)E_{K^+}F_{K^+}\nn
& &+F^{u}_{FF}(Q^2)F_{K^+}F_{K^+})V^{uu}_T(Q^2)\,,
\end{eqnarray}
where $E_{K^{+}},F_{K^{+}}$ is kaon's amplitude and the corresponding functions can be found by the new regularization approach as
\begin{eqnarray}\label{eqn::fueeefff}
F^{u}_{EE}(Q^2)&=&2N_c\left(\int_{x}I_0(\omega_{K2})-\int_{xy}A_{EE} I_{-2}(\omega_{K3})\right)\nn
F^{u}_{EF}(Q^2)&=&-2\frac{M_u+M_s}{M_{us}}F^{u}_{EE}(Q^2)+N_c\int_{x}A_{EF}I_0(\omega_{K2})\nn
F^{u}_{FF}(Q^2)&=&-\frac{M_u+M_s}{2M_{us}}F^{u}_{EF}(Q^2)+N_c\int_{x}A_{FF}I_0(\omega_{K2})\nn
               & &+N_c\int_{x}B_{FF}I_0(M_u^2+x(1-x)Q^2)\,,
\end{eqnarray}
where
\begin{eqnarray}
&&A_{EE}=2 ((M_s - M_u)^2 - m_K^2 ) (1 - x - y) + Q^2 (x + y)\,,\nn
&&A_{EF}=\frac{4 (M_s(1-x) + M_u x)}{M_{us}}\,,\nn
&&A_{FF}=\frac{-2(M_s^2(1-x) + M_u^2 x+M_uM_s)}{M^2_{us}}\,,\nn
&&B_{FF}=\frac{4 Q^2 x(1-x)}{M_{us}^2}\,,
\end{eqnarray}
and $\int_{x}=\int_{0}^{1}dx$, $\int_{xy}=\int_{0}^{1}dx\int_{0}^{1-x}dy$, $\omega_{K2}=xM_u^2+(1-x)M_s^2-x(1-x)M_K^2$, $\omega_{K3}=(x+y)M_u^2+(1-x-y)M_s^2-m_K^2(x + y)(1-x-y)+Q^2 x y$.

\begin{figure}[htbp]
\includegraphics[width=8.6cm]{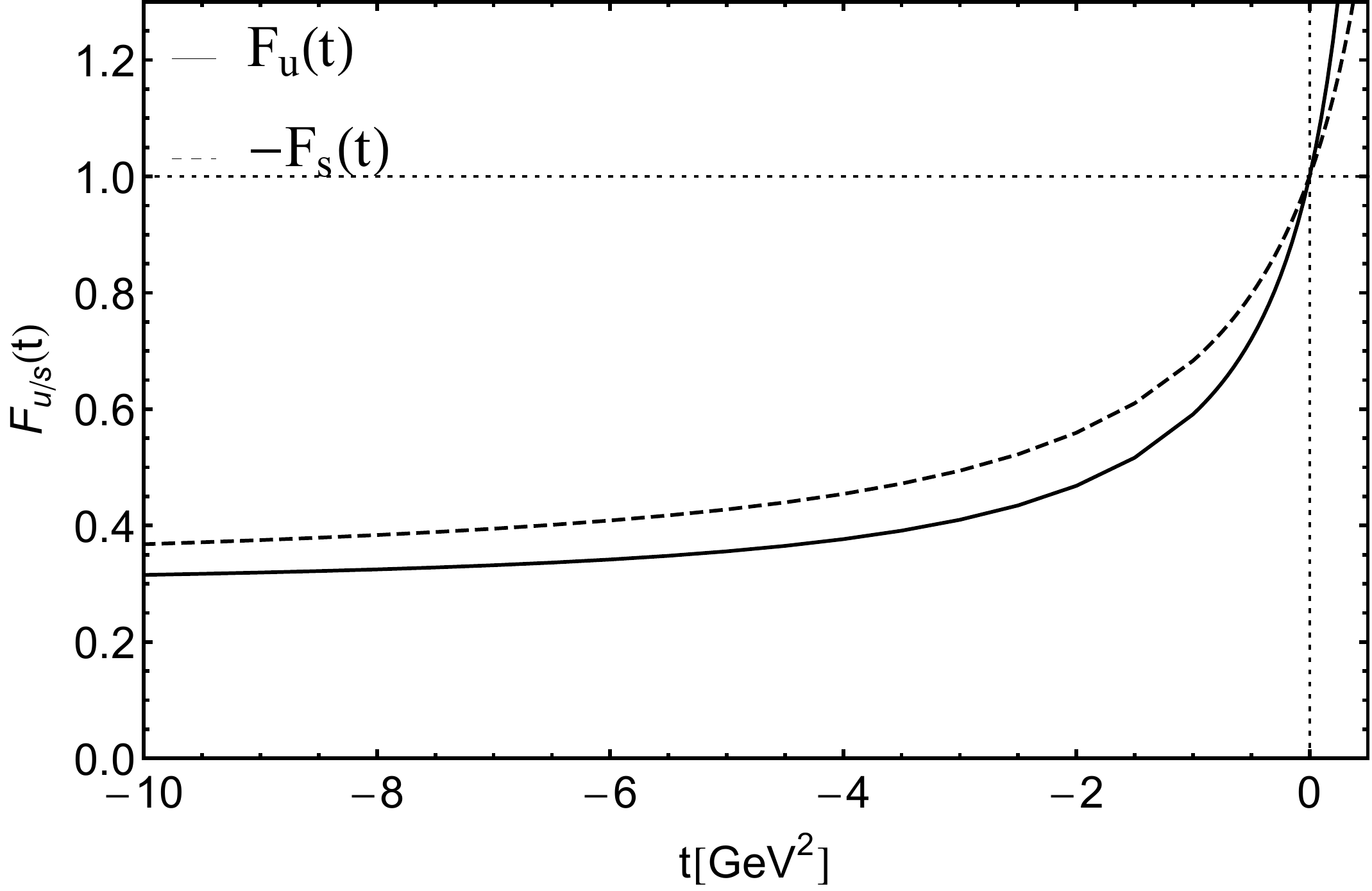}
\caption{The momentum dependence of kaon electromagnetic form factors $F_{u/s}(t=-Q^2)$.}
\label{fig::kem}
\end{figure}

Consider the interchange $s\leftrightarrow u$ and $p\leftrightarrow -k$ in \Eqn{eqn::fs} and comparing it with \Eqn{eqn::fu}, one finds that $F^{s}_{K^+}(Q^2)$ is obtained from $-F^{u}_{K^+}(Q^2)$ by the interchange of $s\leftrightarrow u$. From the explicit expressions of $F^{u}_{EF}(Q^2)$, and some symmetries of the Feynman parameters, one gets the following relation 
\begin{eqnarray}\label{eqn::ffef}
F^{u}_{EF}(Q^2=0)+F^{s}_{EF}(Q^2=0)=0\,.
\end{eqnarray}
Once again we see that the symmetry preserving regularization procedure leads to the desired result, \Eqn{eqn::ffef}, which should always hold, whereas in previous contact interaction studies this relation is broken because of improper regularization procedure. The result for the kaon electromagnetic form factor is presented in \Fig{fig::kem}, which, as expected, is hard due to the nature of contact interaction.

\subsection{$K_{l3}$ transition form factors}
\begin{figure}[htbp]
\includegraphics[width=8.6cm]{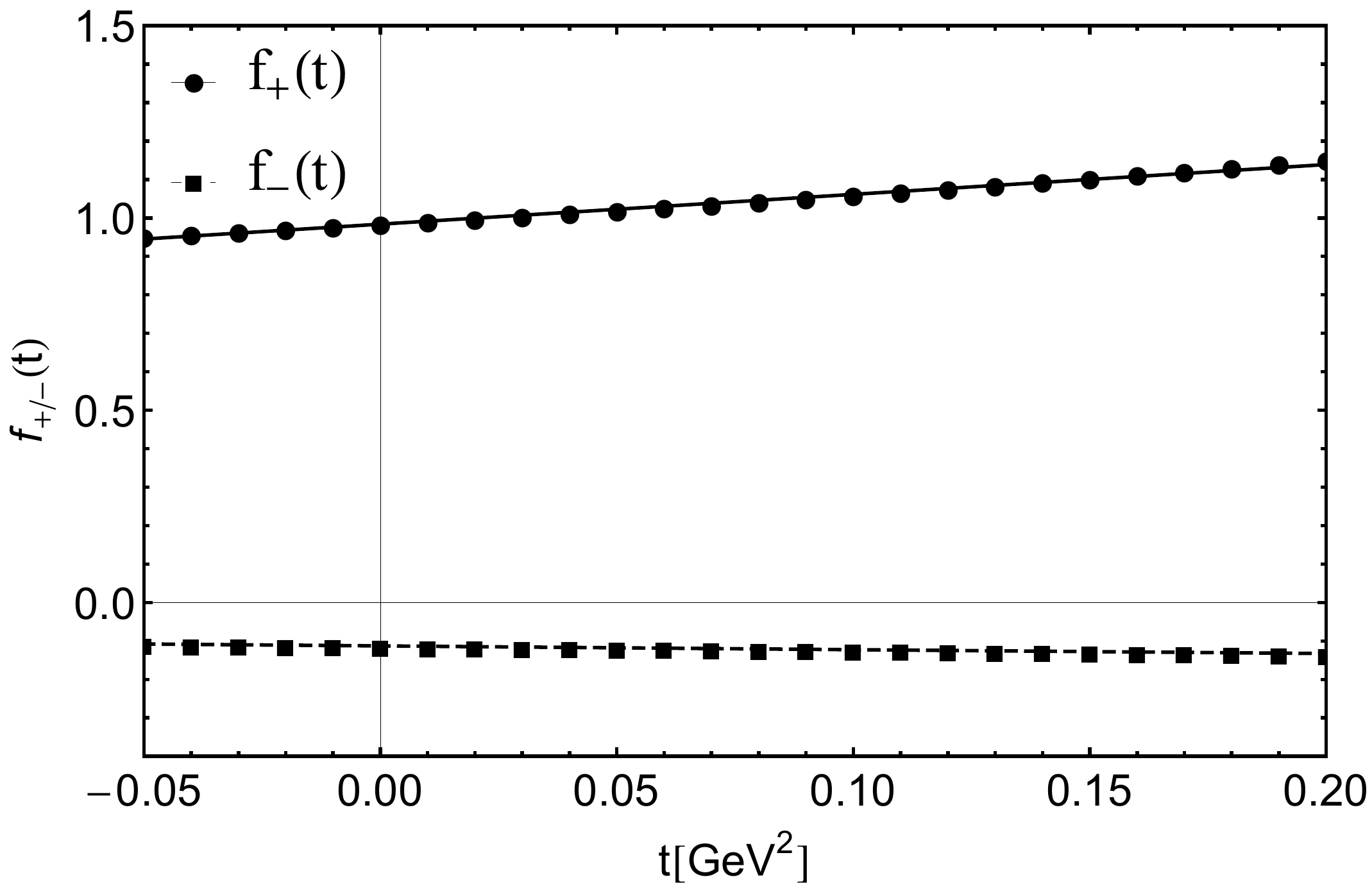}
\includegraphics[width=8.6cm]{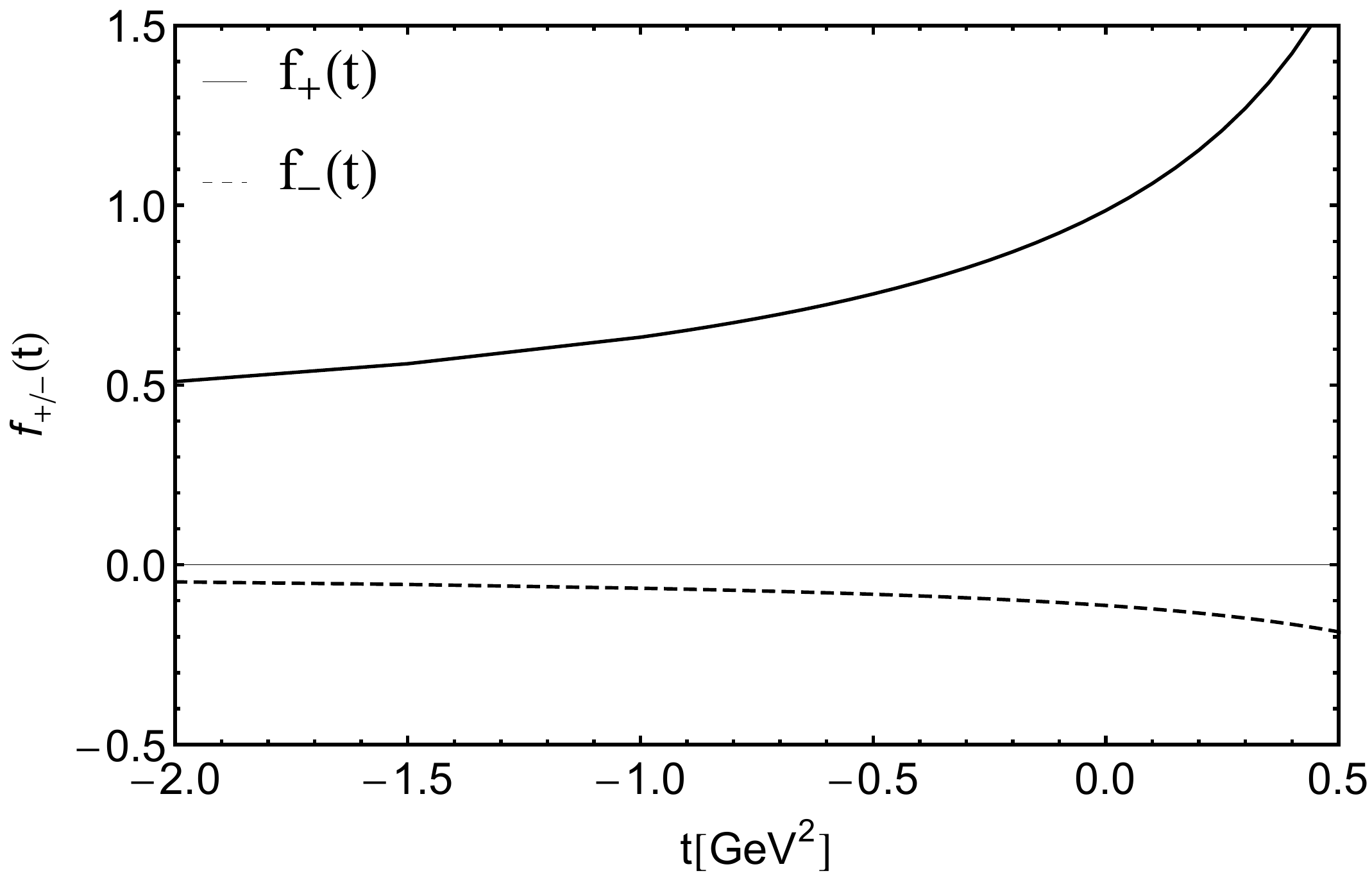}
\captionsetup{justification=raggedright}
\caption{Results for $f_{+/-}(t)$. In upper panel, the filled circles and squares are results of the calculation, the solid and dashed lines are linear fits in \Eqn{lfit}. In lower panel, the solid and dashed lines are the results of the calculation.}
\label{fig::kl3}
\end{figure}

The impulse approximation result for the $K_{l3}$ transition amplitude is
\begin{eqnarray}\label{eqn::kl3amplitude}
M^{K_{l3}}_{\mu}&=&\frac{1}{\sqrt{2}}N_c\text{tr}_{D}\int_q\{S^{u}(q-p)\Gamma^{\pi}(-p)S^{u}(q)\nn
&&\times \Gamma^{K^+}(k)S^{s}(q-k)i\Gamma^{su}_{\mu}(Q)\}\,,
\end{eqnarray}
where $k$ the incoming kaon momentum and $p$ the out going pion momentum. The on-shell conditions entails $P\cdot Q=(m_K^2-m_\pi^2)/2$ and $P^2=-(m_K^2+m_\pi^2)/2-Q^2/4$. 
The amplitude can be expressed as
\begin{equation}
M^{K_{l3}}_{\mu}=\frac{1}{\sqrt{2}}(2P_{\mu}f_{+}(Q^2)-Q_{\mu}f_{-}(Q^2))\,.
\end{equation}
In addition to the primary and secondary transition form factor $f_{+}(Q^2)$ and $f_{-}(Q^2)$, it is helpful to characterize the transitions by  the following function
\begin{equation}
f_{0}(Q^2)=f_{+}(Q^2)-Q^2f_{-}(Q^2)/(m_K^2-m_\pi^2)\,.
\end{equation}
Since the expressions for $f_{+}(Q^2)$ and $f_{-}(Q^2)$ are complicated, we fit these two functions by interpolating in the domain $-2<t/\GeV^2<0.5$.
\begin{eqnarray}
%f_{+}(t)&=&f_{+}(0)\frac{1-0.739735t+0.115904t^2}{1-1.43117t+0.457416t^2}\,,\nn
%f_{-}(t)&=&f_{-}(0)\frac{1-1.29215t+0.0933011t^2}{1-2.07582t+1.06705t^2}\,,
f_{+}(t)&=&f_{+}(0)\frac{1-0.7397t+0.1159t^2}{1-1.4312t+0.4574t^2}\,,\nn
f_{-}(t)&=&f_{-}(0)\frac{1-1.2922t+0.0933t^2}{1-2.0758t+1.0671t^2}\,,
\end{eqnarray}
with $t=-Q^2$ and in the following we use $t$ as variable of $f_{+,-,0}$. 
And in the physical interesting domain $m_l^2<t/\GeV^2<t_m$, where $t_m=(M_K-m_\pi)^2\sim0.13 \GeV^2$, a linear fit is performed,
\begin{eqnarray}\label{lfit}
%f_{+}(t)&=&0.983766 + 0.775075 t\,,\nn
%f_{-}(t)&=&-0.112769 - 0.0991613 t\,.
f_{+}(t)&=&0.9838 + 0.7751 t\,,\nn
-f_{-}(t)&=&0.1128 + 0.0992 t\,.
\end{eqnarray}
The slopes of transition form factors are less deep than the results reported in Ref.~\cite{Ji:2001pj}. The results for the FFs $f_{+}$ and $f_{-}$ are plotted in \Fig{fig::kl3}.

Here we list in Table.~\ref{tab:kaon} a range of quantities that used to characterise the kaon semileptonic decays. The difference between the results of previous work~\cite{Chen:2012txa} and the present results comes from the regularization procedure.

\begin{table}[ht]
%\vspace*{-3mm}
\caption{\label{tab:kaon} quantities derived from the kaon transition form factors}
\begin{tabular}{c|ccccc}
\hline
&$f_{+}(0)$ &$f_{+}(t_m)$ &$-f_{-}(0)$ & $-f_{-}(t_m)$ & $f_{0}(t_\Delta)$ \\
\hline
here&0.98 &1.09 & 0.11 & 0.12 & 1.05\\
\hline
Ref.~\cite{Chen:2012txa}&0.98 &1.07 & 0.087 & 0.096 & 1.06\\
\hline
Ref.~\cite{Ji:2001pj}&0.96 &1.13 & 0.10 & 0.11 & 1.18\\
\hline
Ref.~\cite{Sciascia:2008fr,ParticleDataGroup:2012pjm,Cirigliano:2011ny}&0.96 &1.16 &  0.12 &  &  \\
\hline
\end{tabular}
%\vspace*{-3mm}
\end{table}

Current algebra predicts\cite{Callan:1966hu} that
\begin{equation}
f_{0}(t_\Delta=m_K^2-m_\pi^2)=f_K/f_\pi+\Delta_{CT}\,,
\end{equation}
where $f_{\pi,K}$ is the leptonic decay constant of pion and kaon, and the correction $\Delta_{CT}$ is of $\mathcal{O}(m_u,m_d)$, which is usually negligible. In our calculations, we obtain $\Delta_{CT}/f_{0}(t_\Delta)=3.7\times 10^{-4}$. Actually, $\Delta_{CT}=0$ if one works in the chiral limit, i.e., $m_u=0=m_{\pi}$. It is known that in the solution of \Eqn{eqn::avbse} has the form in the neighbourhood of $P^2=0$(with $a,b=u$)
\begin{equation}\label{eqn::avsolution}
\Gamma_{5\mu}^{uu}(P)=\frac{P_\mu}{P^2}\sqrt{2}f_{\pi}\Gamma^{\pi}(P)+\gamma_5\gamma_\mu F_{R}(P)\,,
\end{equation}
where $F_{R}$ is the regular part of vertex at $P^2=0$.
Plugging \Eqn{eqn::avsolution} into the axial vector WTI \Eqn{eqn::avwti}, one obtains the following generalised Goldberger-Treiman relations\cite{Maris:1997hd}
\begin{eqnarray}
&&\sqrt{2}E_\pi(P) f_\pi=2M_u\,,\\
&&\frac{2F_\pi(P)}{E_\pi(P)}+F_R(P)=1\,.
\end{eqnarray}
Then pion BSA can be expressed as follows
\begin{equation}\label{eqn::chipibsa}
\Gamma^{\pi}(P)=\frac{S^{-1}(q+P)i\gamma_5+i\gamma_5S^{-1}(q)-\gamma_5\slashed{P} F_{R}(P)}{\sqrt{2}f_{\pi}}\,.
\end{equation}
Substitute \Eqn{eqn::chipibsa} into the $K_{l3}$ transition amplitude \Eqn{eqn::kl3amplitude}, one arrives at the following expressions for the transition form factors
\begin{eqnarray}
f^{cl}_{+}(t)&=&\frac{f_K}{2f_\pi}-\frac{E_\pi-2F_\pi}{2}(f_{cl+}^{FE}(t)E_{K}+f_{cl+}^{FF}(t)F_{K})\,,\nn
f^{cl}_{-}(t)&=&\frac{f_K}{2f_\pi}-\frac{E_\pi-2F_\pi}{2}(f_{cl-}^{FE}(t)E_{K}+f_{cl-}^{FF}(t)F_{K})\,,\nn
\end{eqnarray}
where $f_{+/-}^{FX}(t)$ with $X=E,F$ comes from the following way of expressing the transition form factor in analogy with \Eqn{eqn::emff}
\begin{eqnarray}
f_{+/-}(t)&=&(f_{+/-}^{EE}(t)E_{\pi}E_{K}+f_{+/-}^{EF}(t)E_{\pi}F_{K}\nn
& &+f_{+/-}^{FE}(t)F_{\pi}E_{K}+f_{+/-}^{FF}(t)F_{\pi}F_{K})\,,
\end{eqnarray}
and $f^{FX}_{cl+/-}(t)=f^{FX}_{+/-}(t,m_\pi=0)$. Furthermore, one can extract the vector vertex ingredients,
\begin{eqnarray}
f_{+}^{FX}(t)&=&f_{+}^{FX,T}(t)V_T^{su}(t)\,,\\
f_{-}^{FX}(t)&=&f_{-}^{FX,T}(t)V_T^{su}(t)+f_{-}^{FX,L}(t)V_L^{su}(t)\nn
& &+f_{-}^{FX,S}(t)V_S^{su}(t)\,.
\end{eqnarray}
Our computation indicates the following relations in the chiral limit
\begin{eqnarray}
f_{cl+}^{FX,T}(t)+\frac{t}{m_K^2}f_{c-}^{FX,T}(t)=0\,,\\
f_{cl-}^{FX,L}(t)\sim \mathcal{O}(t-m_K^2)\,,\\
f_{cl-}^{FX,S}(t)\sim \mathcal{O}(t-m_K^2)\,,
\end{eqnarray}
which leads to
\begin{eqnarray}
f^{cl}_{0}(t_\Delta)=f^{cl}_{+}(t_\Delta)+f^{cl}_{-}(t_\Delta)=\frac{f_K}{f_{\pi}}\,.
\end{eqnarray}
Hence $\Delta_{CT}$ vanishes in the chiral limit.

\section{summary}\label{sec::summary}
In this paper, we introduce a symmetry preserving regularization based on Schwinger's proper-time method. This regularization meets many good features of dimensional regularization without changing space-time dimensions. One of the most important properties is that it ensures the regularized ILIs satisfy a series of consistency conditions which are necessary to preserving gauge symmetries. In particular, we show that gauge symmetries are preserved by proving that the WTIs hold exactly after the regularization. We also show why the regularization procedure in previous studies of the contact model breaks WTIs. Systematic steps for regularizing the contact model are presented at the end of Sec.~\ref{sec::contactmodel}. 

As an application example, we recalculated kaon form factors which exhibits inconsistency in Ref.~\cite{Chen:2012txa} because of the improperly regularized ILIs, whereas under the new regularization the results are self-consistent. The present regularization would also rescue the flaw in the calculation of other form factors, such as heavy-light mesons semileptonic transitions(see ~\cite{Xu:2021iwv} for some discussion).

With this symmetry preserving regularization, the hadron properties computed from contact model are valid, as potential inconsistencies from regularization are eliminated. It is hope that the contact model studies can provide more inspiring results that may shed lights on the realistic DSE studies.

\section{acknowledgements}
Work supported by National Natural Science Foundation of China (grant no. 12135007).

\bibliography{apstemplateNotes}

\providecommand{\noopsort}[1]{}\providecommand{\singleletter}[1]{#1}%
\begin{thebibliography}{45}
\providecommand{\natexlab}[1]{#1}
\providecommand{\url}[1]{\texttt{#1}}
\expandafter\ifx\csname urlstyle\endcsname\relax
  \providecommand{\doi}[1]{doi: #1}\else
  \providecommand{\doi}{doi: \begingroup \urlstyle{rm}\Url}\fi

\bibitem[Roberts and Williams(1994)]{Roberts:1994dr}
Craig~D. Roberts and Anthony~G. Williams.
\newblock {Dyson-Schwinger equations and their application to hadronic
  physics}.
\newblock \emph{Prog. Part. Nucl. Phys.}, 33:\penalty0 477--575, 1994.
\newblock \doi{10.1016/0146-6410(94)90049-3}.

\bibitem[Maris and Roberts(2003)]{Maris:2003vk}
Pieter Maris and Craig~D. Roberts.
\newblock {Dyson-Schwinger equations: A Tool for hadron physics}.
\newblock \emph{Int. J. Mod. Phys. E}, 12:\penalty0 297--365, 2003.
\newblock \doi{10.1142/S0218301303001326}.

\bibitem[Fischer(2019)]{Fischer:2018sdj}
Christian~S. Fischer.
\newblock {QCD at finite temperature and chemical potential from
  Dyson\textendash{}Schwinger equations}.
\newblock \emph{Prog. Part. Nucl. Phys.}, 105:\penalty0 1--60, 2019.
\newblock \doi{10.1016/j.ppnp.2019.01.002}.

\bibitem[Eichmann(2009)]{Eichmann:2009zx}
Gernot Eichmann.
\newblock \emph{{Hadron properties from QCD bound-state equations}}.
\newblock PhD thesis, Graz U., 2009.

\bibitem[Eichmann et~al.(2016)Eichmann, Sanchis-Alepuz, Williams, Alkofer, and
  Fischer]{Eichmann:2016yit}
Gernot Eichmann, Helios Sanchis-Alepuz, Richard Williams, Reinhard Alkofer, and
  Christian~S. Fischer.
\newblock {Baryons as relativistic three-quark bound states}.
\newblock \emph{Prog. Part. Nucl. Phys.}, 91:\penalty0 1--100, 2016.
\newblock \doi{10.1016/j.ppnp.2016.07.001}.

\bibitem[Gutierrez-Guerrero et~al.(2010)Gutierrez-Guerrero, Bashir, Cloet, and
  Roberts]{Gutierrez-Guerrero:2010waf}
L.~X. Gutierrez-Guerrero, A.~Bashir, I.~C. Cloet, and C.~D. Roberts.
\newblock {Pion form factor from a contact interaction}.
\newblock \emph{Phys. Rev. C}, 81:\penalty0 065202, 2010.
\newblock \doi{10.1103/PhysRevC.81.065202}.

\bibitem[Roberts et~al.(2011{\natexlab{a}})Roberts, Chang, Cloet, and
  Roberts]{Roberts:2011cf}
Hannes L.~L. Roberts, Lei Chang, Ian~C. Cloet, and Craig~D. Roberts.
\newblock {Masses of ground and excited-state hadrons}.
\newblock \emph{Few Body Syst.}, 51:\penalty0 1--25, 2011{\natexlab{a}}.
\newblock \doi{10.1007/s00601-011-0225-x}.

\bibitem[Roberts et~al.(2011{\natexlab{b}})Roberts, Bashir, Gutierrez-Guerrero,
  Roberts, and Wilson]{Roberts:2011wy}
H.~L.~L. Roberts, A.~Bashir, L.~X. Gutierrez-Guerrero, C.~D. Roberts, and D.~J.
  Wilson.
\newblock {pi- and rho-mesons, and their diquark partners, from a contact
  interaction}.
\newblock \emph{Phys. Rev. C}, 83:\penalty0 065206, 2011{\natexlab{b}}.
\newblock \doi{10.1103/PhysRevC.83.065206}.

\bibitem[Chen et~al.(2012)Chen, Chang, Roberts, Wan, and Wilson]{Chen:2012qr}
Chen Chen, Lei Chang, Craig~D. Roberts, Shaolong Wan, and David~J. Wilson.
\newblock {Spectrum of hadrons with strangeness}.
\newblock \emph{Few Body Syst.}, 53:\penalty0 293--326, 2012.
\newblock \doi{10.1007/s00601-012-0466-3}.

\bibitem[Chen et~al.(2013)Chen, Chang, Roberts, Schmidt, Wan, and
  Wilson]{Chen:2012txa}
Chen Chen, Lei Chang, Craig~D. Roberts, Sebastian~M. Schmidt, Shaolong Wan, and
  David~J. Wilson.
\newblock {Features and flaws of a contact interaction treatment of the kaon}.
\newblock \emph{Phys. Rev. C}, 87:\penalty0 045207, 2013.
\newblock \doi{10.1103/PhysRevC.87.045207}.

\bibitem[Segovia et~al.(2014{\natexlab{a}})Segovia, Chen, Clo\"et, Roberts,
  Schmidt, and Wan]{Segovia:2013uga}
Jorge Segovia, Chen Chen, Ian~C. Clo\"et, Craig~D. Roberts, Sebastian~M.
  Schmidt, and Shaolong Wan.
\newblock {Elastic and Transition Form Factors of the $\Delta(1232)$}.
\newblock \emph{Few Body Syst.}, 55:\penalty0 1--33, 2014{\natexlab{a}}.
\newblock \doi{10.1007/s00601-013-0734-x}.

\bibitem[Segovia et~al.(2014{\natexlab{b}})Segovia, Cloet, Roberts, and
  Schmidt]{Segovia:2014aza}
Jorge Segovia, Ian~C. Cloet, Craig~D. Roberts, and Sebastian~M. Schmidt.
\newblock {Nucleon and $\Delta$ elastic and transition form factors}.
\newblock \emph{Few Body Syst.}, 55:\penalty0 1185--1222, 2014{\natexlab{b}}.
\newblock \doi{10.1007/s00601-014-0907-2}.

\bibitem[Xu et~al.(2015)Xu, Chen, Cloet, Roberts, Segovia, and
  Zong]{Xu:2015kta}
Shu-Sheng Xu, Chen Chen, Ian~C. Cloet, Craig~D. Roberts, Jorge Segovia, and
  Hong-Shi Zong.
\newblock {Contact-interaction Faddeev equation and, inter alia , proton tensor
  charges}.
\newblock \emph{Phys. Rev. D}, 92\penalty0 (11):\penalty0 114034, 2015.
\newblock \doi{10.1103/PhysRevD.92.114034}.

\bibitem[Bedolla et~al.(2015)Bedolla, Cobos-Mart\'\i{}nez, and
  Bashir]{Bedolla:2015mpa}
Marco~A. Bedolla, J.~J. Cobos-Mart\'\i{}nez, and Adnan Bashir.
\newblock {Charmonia in a contact interaction}.
\newblock \emph{Phys. Rev. D}, 92\penalty0 (5):\penalty0 054031, 2015.
\newblock \doi{10.1103/PhysRevD.92.054031}.

\bibitem[Serna et~al.(2016)Serna, Brito, and Krein]{Serna:2016kdb}
F.~E. Serna, M.~A. Brito, and G.~Krein.
\newblock {Symmetry-preserving contact interaction model for heavy-light
  mesons}.
\newblock \emph{AIP Conf. Proc.}, 1701\penalty0 (1):\penalty0 100018, 2016.
\newblock \doi{10.1063/1.4938727}.

\bibitem[Lu et~al.(2017)Lu, Chen, Roberts, Segovia, Xu, and Zong]{Lu:2017cln}
Ya~Lu, Chen Chen, Craig~D. Roberts, Jorge Segovia, Shu-Sheng Xu, and Hong-Shi
  Zong.
\newblock {Parity partners in the baryon resonance spectrum}.
\newblock \emph{Phys. Rev. C}, 96\penalty0 (1):\penalty0 015208, 2017.
\newblock \doi{10.1103/PhysRevC.96.015208}.

\bibitem[Serna et~al.(2017)Serna, El-Bennich, and Krein]{Serna:2017nlr}
Fernando~E. Serna, Bruno El-Bennich, and Gast\~ao Krein.
\newblock {Charmed mesons with a symmetry-preserving contact interaction}.
\newblock \emph{Phys. Rev. D}, 96\penalty0 (1):\penalty0 014013, 2017.
\newblock \doi{10.1103/PhysRevD.96.014013}.

\bibitem[Guti\'errez-Guerrero et~al.(2019)Guti\'errez-Guerrero, Bashir,
  Bedolla, and Santopinto]{Gutierrez-Guerrero:2019uwa}
L.~X. Guti\'errez-Guerrero, Adnan Bashir, Marco~A. Bedolla, and E.~Santopinto.
\newblock {Masses of Light and Heavy Mesons and Baryons: A Unified Picture}.
\newblock \emph{Phys. Rev. D}, 100\penalty0 (11):\penalty0 114032, 2019.
\newblock \doi{10.1103/PhysRevD.100.114032}.

\bibitem[Zhang et~al.(2021)Zhang, Cui, Ping, and Roberts]{Zhang:2020ecj}
Jin-Li Zhang, Zhu-Fang Cui, Jialun Ping, and Craig~D Roberts.
\newblock {Contact interaction analysis of pion GTMDs}.
\newblock \emph{Eur. Phys. J. C}, 81\penalty0 (1):\penalty0 6, 2021.
\newblock \doi{10.1140/epjc/s10052-020-08791-1}.

\bibitem[Guti\'errez-Guerrero et~al.(2021)Guti\'errez-Guerrero, Paredes-Torres,
  and Bashir]{Gutierrez-Guerrero:2021rsx}
L.~X. Guti\'errez-Guerrero, G.~Paredes-Torres, and A.~Bashir.
\newblock {Mesons and baryons: Parity partners}.
\newblock \emph{Phys. Rev. D}, 104\penalty0 (9):\penalty0 094013, 2021.
\newblock \doi{10.1103/PhysRevD.104.094013}.

\bibitem[Xing et~al.(2021)Xing, Raya, and Chang]{Xing:2021dwe}
Zanbin Xing, Kh\'epani Raya, and Lei Chang.
\newblock {Quark anomalous magnetic moment and its effects on the
  \ensuremath{\rho} meson properties}.
\newblock \emph{Phys. Rev. D}, 104\penalty0 (5):\penalty0 054038, 2021.
\newblock \doi{10.1103/PhysRevD.104.054038}.

\bibitem[Wu(2003)]{Wu:2002xa}
Yue-Liang Wu.
\newblock {Symmetry principle preserving and infinity free regularization and
  renormalization of quantum field theories and the mass gap}.
\newblock \emph{Int. J. Mod. Phys. A}, 18:\penalty0 5363--5420, 2003.
\newblock \doi{10.1142/S0217751X03015222}.

\bibitem[Wu(2004)]{Wu:2003dd}
Yue-Liang Wu.
\newblock {Symmetry preserving loop regularization and renormalization of
  QFTs}.
\newblock \emph{Mod. Phys. Lett. A}, 19:\penalty0 2191--2204, 2004.
\newblock \doi{10.1142/S0217732304015361}.

\bibitem[Ebert et~al.(1996)Ebert, Feldmann, and Reinhardt]{Ebert:1996vx}
Dietmar Ebert, Thorsten Feldmann, and Hugo Reinhardt.
\newblock {Extended NJL model for light and heavy mesons without q - anti-q
  thresholds}.
\newblock \emph{Phys. Lett. B}, 388:\penalty0 154--160, 1996.
\newblock \doi{10.1016/0370-2693(96)01158-6}.

\bibitem[Rochester and Butler(1947)]{Rochester:1947mi}
G.~D. Rochester and C.~C. Butler.
\newblock {Evidence for the Existence of New Unstable Elementary Particles}.
\newblock \emph{Nature}, 160:\penalty0 855--857, 1947.
\newblock \doi{10.1038/160855a0}.

\bibitem[Gell-Mann(1953)]{Gell-Mann:1953hzm}
M.~Gell-Mann.
\newblock {Isotopic Spin and New Unstable Particles}.
\newblock \emph{Phys. Rev.}, 92:\penalty0 833--834, 1953.
\newblock \doi{10.1103/PhysRev.92.833}.

\bibitem[Lee and Yang(1956)]{Lee:1956qn}
T.~D. Lee and Chen-Ning Yang.
\newblock {Question of Parity Conservation in Weak Interactions}.
\newblock \emph{Phys. Rev.}, 104:\penalty0 254--258, 1956.
\newblock \doi{10.1103/PhysRev.104.254}.

\bibitem[Cabibbo(1963)]{Cabibbo:1963yz}
Nicola Cabibbo.
\newblock {Unitary Symmetry and Leptonic Decays}.
\newblock \emph{Phys. Rev. Lett.}, 10:\penalty0 531--533, 1963.
\newblock \doi{10.1103/PhysRevLett.10.531}.

\bibitem[Kobayashi and Maskawa(1973)]{Kobayashi:1973fv}
Makoto Kobayashi and Toshihide Maskawa.
\newblock {CP Violation in the Renormalizable Theory of Weak Interaction}.
\newblock \emph{Prog. Theor. Phys.}, 49:\penalty0 652--657, 1973.
\newblock \doi{10.1143/PTP.49.652}.

\bibitem[Christenson et~al.(1964)Christenson, Cronin, Fitch, and
  Turlay]{Christenson:1964fg}
J.~H. Christenson, J.~W. Cronin, V.~L. Fitch, and R.~Turlay.
\newblock {Evidence for the $2\pi$ Decay of the $K_2^0$ Meson}.
\newblock \emph{Phys. Rev. Lett.}, 13:\penalty0 138--140, 1964.
\newblock \doi{10.1103/PhysRevLett.13.138}.

\bibitem[Glashow et~al.(1970)Glashow, Iliopoulos, and Maiani]{Glashow:1970gm}
S.~L. Glashow, J.~Iliopoulos, and L.~Maiani.
\newblock {Weak Interactions with Lepton-Hadron Symmetry}.
\newblock \emph{Phys. Rev. D}, 2:\penalty0 1285--1292, 1970.
\newblock \doi{10.1103/PhysRevD.2.1285}.

\bibitem[Taboada-Nieto et~al.(2022)Taboada-Nieto, Ortega, Entem, Fern\'andez,
  and Segovia]{Taboada-Nieto:2022igy}
U.~Taboada-Nieto, P.~G. Ortega, D.~R. Entem, F.~Fern\'andez, and J.~Segovia.
\newblock {Kaon spectrum revisited}.
\newblock 9 2022.

\bibitem[Maris and Tandy(2000)]{Maris:2000sk}
Pieter Maris and Peter~C. Tandy.
\newblock {The pi, K+, and K0 electromagnetic form-factors}.
\newblock \emph{Phys. Rev. C}, 62:\penalty0 055204, 2000.
\newblock \doi{10.1103/PhysRevC.62.055204}.

\bibitem[Xiao et~al.(2002)Xiao, Qian, and Ma]{Xiao:2002iv}
Bo-Wen Xiao, Xin Qian, and Bo-Qiang Ma.
\newblock {The Kaon form-factor in the light cone quark model}.
\newblock \emph{Eur. Phys. J. A}, 15:\penalty0 523--527, 2002.
\newblock \doi{10.1140/epja/i2002-10059-y}.

\bibitem[Stamen et~al.(2022)Stamen, Hariharan, Hoferichter, Kubis, and
  Stoffer]{Stamen:2022uqh}
Dominik Stamen, Deepti Hariharan, Martin Hoferichter, Bastian Kubis, and Peter
  Stoffer.
\newblock {Kaon electromagnetic form factors in dispersion theory}.
\newblock \emph{Eur. Phys. J. C}, 82\penalty0 (5):\penalty0 432, 2022.
\newblock \doi{10.1140/epjc/s10052-022-10348-3}.

\bibitem[Ji and Maris(2001)]{Ji:2001pj}
Chueng-Ryong Ji and Pieter Maris.
\newblock {K(l3) transition form-factors}.
\newblock \emph{Phys. Rev. D}, 64:\penalty0 014032, 2001.
\newblock \doi{10.1103/PhysRevD.64.014032}.

\bibitem[Bazavov et~al.(2013)]{Bazavov:2012cd}
A.~Bazavov et~al.
\newblock {Kaon semileptonic vector form factor and determination of $|V_{us}|$
  using staggered fermions}.
\newblock \emph{Phys. Rev. D}, 87:\penalty0 073012, 2013.
\newblock \doi{10.1103/PhysRevD.87.073012}.

\bibitem[Boyle et~al.(2015)]{RBCUKQCD:2015joy}
Peter~A. Boyle et~al.
\newblock {The kaon semileptonic form factor in N$_{f}$ = 2 + 1 domain wall
  lattice QCD with physical light quark masses}.
\newblock \emph{JHEP}, 06:\penalty0 164, 2015.
\newblock \doi{10.1007/JHEP06(2015)164}.

\bibitem[Troitsky and Troitsky(2021)]{Troitsky:2021vkw}
S.~V. Troitsky and V.~E. Troitsky.
\newblock {K0 and K+-meson electromagnetic form factors: A nonperturbative
  relativistic quark model versus experimental, perturbative, and lattice
  quantum-chromodynamics results}.
\newblock \emph{Phys. Rev. D}, 104\penalty0 (3):\penalty0 034015, 2021.
\newblock \doi{10.1103/PhysRevD.104.034015}.

\bibitem[Sciascia(2008)]{Sciascia:2008fr}
B.~Sciascia.
\newblock {Precision tests of the SM with leptonic and semileptonic kaon
  decays}.
\newblock \emph{Nucl. Phys. B Proc. Suppl.}, 181-182:\penalty0 83--88, 2008.
\newblock \doi{10.1016/j.nuclphysbps.2008.09.008}.

\bibitem[Beringer et~al.(2012)]{ParticleDataGroup:2012pjm}
J.~Beringer et~al.
\newblock {Review of Particle Physics (RPP)}.
\newblock \emph{Phys. Rev. D}, 86:\penalty0 010001, 2012.
\newblock \doi{10.1103/PhysRevD.86.010001}.

\bibitem[Cirigliano et~al.(2012)Cirigliano, Ecker, Neufeld, Pich, and
  Portoles]{Cirigliano:2011ny}
Vincenzo Cirigliano, Gerhard Ecker, Helmut Neufeld, Antonio Pich, and Jorge
  Portoles.
\newblock {Kaon Decays in the Standard Model}.
\newblock \emph{Rev. Mod. Phys.}, 84:\penalty0 399, 2012.
\newblock \doi{10.1103/RevModPhys.84.399}.

\bibitem[Callan and Treiman(1966)]{Callan:1966hu}
C.~G. Callan and S.~B. Treiman.
\newblock {Equal Time Commutators and K Meson Decays}.
\newblock \emph{Phys. Rev. Lett.}, 16:\penalty0 153--157, 1966.
\newblock \doi{10.1103/PhysRevLett.16.153}.

\bibitem[Maris et~al.(1998)Maris, Roberts, and Tandy]{Maris:1997hd}
Pieter Maris, Craig~D. Roberts, and Peter~C. Tandy.
\newblock {Pion mass and decay constant}.
\newblock \emph{Phys. Lett. B}, 420:\penalty0 267--273, 1998.
\newblock \doi{10.1016/S0370-2693(97)01535-9}.

\bibitem[Xu et~al.(2021)Xu, Cui, Roberts, and Xu]{Xu:2021iwv}
Zhen-Ni Xu, Zhu-Fang Cui, Craig~D. Roberts, and Chang Xu.
\newblock {Heavy + light pseudoscalar meson semileptonic transitions}.
\newblock \emph{Eur. Phys. J. C}, 81\penalty0 (12):\penalty0 1105, 2021.
\newblock \doi{10.1140/epjc/s10052-021-09898-9}.

\end{thebibliography}
\end{document}